\documentstyle[12pt,aasms4]{article}

\received{28 Febraury 2001}
\accepted{28 Febraury 2001}
\slugcomment{to appear in "Emission Lines From Jet Flows", Rev.Mex.Astron. and Astrofis. Conf. Ser.}

\lefthead{Rosado, de la Fuente, Arias and Le Coarer.}
\righthead{A Bipolar Outflow in the Orion nebula}

\begin{document}
\title{Fabry-Perot Kinematics of HH 202, 203-204 in the Orion Nebula: Are they
part of a Big Bipolar Outflow?} \author{ Margarita Rosado$^1$, Eduardo de
la Fuente$^{1,2}$, Lorena Arias $^1$ and  E. Le Coarer
$^3$}  \altaffiltext{1}{Instituto de Astronom\'{\i}a, UNAM: margarit, edfuente, lorena@astroscu.unam.mx}
\altaffiltext{2}{Facultad de Ciencias, UNAM} \altaffiltext{3}{ Observatoire de
Grenoble: lecoarer@obs.ujf-grenoble.fr}

\begin{abstract}
 
We present a kinematic study of the Herbig-Haro objects HH
202, 203 and 204 using  H$\alpha$ and [NII] Fabry-Perot velocity maps. For HH 202
we find new features that could belong to this HH object or that perhaps are associated with an outflow different from HH 202. Because of its high velocity (up to 100 km s$^{-1}$) this outflow probably can be a HH flow not catalogued previously. Large internal motions are found in the fainter regions of  HH 203-204, as well as evidence of transverse density gradients. We show that the apex of HH 204 is the zone of maximum velocity in agreement with bow shock models. 
From our studies, we find kinematic evidence that suggests that HH 203-204 and HH 202 are part of a single and large ($\sim$ 0.55 pc)  HH flow. 
Se presenta un estudio cinem\'atico de los objetos HH 202, 203 y 204
usando mapas de velocidad Fabry-Perot en H$\alpha$ y [NII]. En el caso de HH
202 se encuentran nebulosidades que podr\'{\i}an estar asociadas a este
objeto o bien, dada sus altas velocidades (superiores a 100 km s$^{-1}$), 
formar un flujo HH a\'un no catalogado. Se encuentran movimientos internos
violentos en las regiones d\'ebiles de HH 203 y HH 204 al igual que cierta
evidencia sobre la existencia de gradientes de densidad transversos. Se
muestra que el \'apex de HH 204 corresponde a la zona de velocidad m\'axima,
lo cual est\'a en acuerdo con los modelos de choque de proa. Nuestros estudios
cinem\'aticos nos hacen sugerir que HH 202 y HH 203-204 son parte de un
\'unico  flujo HH bipolar gigante ($\sim$ 0.55 pc). 

\end{abstract}



\section{Introduction}

 The Orion Nebula is one of the most interesting HII regions that still reveals intriguing phenomena. Indeed, Orion  is one of the nearest star forming regions (located at a distance of 450 pc) and, consequently the phenomena can be studied with the highest angular resolution. In the Orion region coexist several phases of the interstellar medium, such as molecular, neutral, and ionized gas. With the development of near IR imaging and spectroscopy in recent years,  new phenomena have been revealed that constitute important clues to the understanding of the process of star formation and the evolution of shocks in a molecular environment. In this work we show the results of a Fabry-Perot kinematical study of the Orion Nebula that was mainly motivated  by: 

-- The excellent HST image coverage  of the Orion Nebula that reveals a plethora of new objects like Herbig-Haro (HH) objects and proplyds (O'Dell {\it {et al}}. 1997a).
 
-- The discovery of giant ($\sim$1 pc) HH flows that challenge our ideas of the energetics and timescales involved in the star formation process (Reipurth {\it {et al}}. 1997). These giant HH flows are constituted of
two or more HH objects that, in the past, were thought to be isolated.

-- The interest in studying jets which are photo-ionized by external sources such as the jet discovered in the Trifid Nebula (Cernicharo {\it {et al}}. 1997). This kind of jet have complex emission and kinematical properties which are now being studied theoretically.

With the aim of obtaining a global view on the kinematics of HH objects and jets in Orion that allows us to search for large scale features that link  the HH objects already known, we undertook the study of the inner 5' of the Orion Nebula by means of Fabry-Perot observations at H$\alpha$ and [NII]($\lambda$ 6583 \AA).
Here we present some of the results derived from this study.

\section{Observations and data reduction}

The Fabry-Perot (FP) observations were carried out  at the f/7.5 Cassegrain focus of the 2.1 m
telescope of the Observatorio Astron\'omico Nacional at San Pedro M\'artir,
B.C. (M\'exico) using the UNAM Scanning Fabry-Perot Interferometer PUMA
(Rosado {\it {et al}}. 1995). We used a 1024$\times$1024 thinned Tektronix CCD
detector, with an image scale of 0.59 arcsec~pixel$^{-1}$.  The interference
filters used in the observations are centered at  H$\alpha$ and [NII]($\lambda$ 
6583 \AA) with bandpasses of 20 and 10 \AA, respectively.  
The main characteristics
of this interferometer are: interference order of 330, free spectral range of
19.89 \AA\ (equivalent to a velocity range of 908
km~s$^{-1}$) and sampling spectral resolution of 0.41 \AA\
(equivalent to 18.9 km~s$^{-1}$), at H$\alpha$, achieved by 
scanning the interferometer gap at 48 positions. Thus, the resulting data
cubes have dimensions of 512$\times$512$\times$48.

With this setup, we
have obtained two nebular data cubes at H$\alpha$ and [NII] with  total  exposure times of 48
and 144 min.,  respectively. 
The data reduction and analysis were performed using the data reduction package CIGALE (Le Coarer {\it {et al}}. 1993).

\section{Kinematical Results}

Figure 1  shows the velocity map at $V_{helio}$ = --127 km
s$^{-1}$ (i.e., ~ 150 km s$^{-1}$ blueshifted relative to the velocity of the intense nebular
background) at H$\alpha$ in which we have
identified the different objects present in the inner 5' region of the Orion Nebula: the HH objects HH 202, HH 269
(barely seen at H$\alpha$), HH 203 and HH 204, the E-W jet mentioned by O'Dell {\it {et al}}. (1997b). In what follows, we will describe some of our results on the kinematics of the HH objects HH 202, HH 203 and HH 204 and discuss the possible existence of a large HH bipolar outflow.

\subsection{ HH 202}

HH 202 was discovered by Cant\'o {\it {et al}}. (1980) as an emission line object showing two knots  embedded in an arc-shaped nebulosity.   Herbig \& Jones (1981) obtained proper
motions for the knots, finding that the proper motion vectors were
parallel and pointing towards the concave side of the curved shell and that there was a considerable dispersion in the tangential speeds of
 several of the knots: from 100 to 294 km s$^{-1}$.  More recently, O'Dell {\it {et
al}}. (1997a) have obtained wonderful HST  images of the HH 202 region in the
[SII], H$\alpha$ and [OIII] lines showing that the arc-shaped nebulosity is
visible in [SII], H$\alpha$ and [OIII] lines. O'Dell
{\it {et al}}. (1997b) obtained [SII] and [OIII] FP spectroscopy of the 
Trapezium region. They found a
blueshifted portion of HH 202 that extends towards the NW. 

From our FP work we see that HH 202 has a different morphology in H$\alpha$ and in [NII]. Figure 2 shows a close-up  of the [NII] velocity map at 
$V_{helio}$ = -- 90 km s$^{-1}$  showing the field near to HH 202 which is
located at the NW corner. Comparing with the H$\alpha$ velocity map (shown in Figure 1) we see that the HH 202 object itself has  different morphologies at 
H$\alpha$ and [NII]. Indeed, while in [NII] we see the 
arc-shaped nebulosity ending to the SW in the southern knot, in H$\alpha$ we detect,  the
arc-shaped nebulosity, but the southern knot is not easily disentangled
from the emission of a bright  nebulosity and several knots. Instead, we see  a bright head of irregular shape
(somewhat like an arrow head) pointing in the E-W  direction,  with three faint
filamentary extensions also oriented in the E-W direction. The longest
filamentary extensions seem to form a rotated spur, or $\Omega$.

 Another interesting feature revealed in Figures 1 and 2 is that HH 202 seems to be part of a larger scale lobe (of 82" $\times$ 25" and aperture angle of  $\sim$ 40 \arcdeg),  oriented in the  NW-SE direction, ending at the position of HH 202 on one side, and
in a point close to the E-W jet in its SE end. The [NII] emission shows that, inside this lobe, two elongated
cavities (one of 35" $\times$ 9" and other of 55" $\times$ 12", both with  aperture angles of  $\sim$ 20 \arcdeg), somewhat similar to bow
shocks  are detected. In H$\alpha$, the walls of the lobe are clearly seen but,
the elongated cavities interior to the lobe (hereafter called `fingers'), are confused with the bright background nebula.

Our FP data show that  high internal motions of the known regions of HH 202  reach blueshifted velocities of up to 100
km s$^{-1}$. The spur also shows this range of velocities
indicating that it should belong to HH 202. On the other hand, the 
lobe and the `fingers' inside it also have  high blueshifted velocities
reaching  up to 100 km s$^{-1}$. This shows that these new features
are related to a high velocity flow, probably an HH flow, not catalogued as
such because of the difficulties of disentangling it from the bright nebular
background. It is unclear whether this flow is associated with
HH 202 or whether it constitutes another HH system.

\subsection{ HH 203 and HH 204}

  These objects were discovered by Munch \& Wilson (1962). Taylor \&
Munch (1978) give radial velocities and velocity
dispersions for several different features within these objects. However, it was Cant\'o {\it {et al}}. (1980) who identified these objects as HH objects.  Hu (1996) has 
measured tangential velocities of 0 and 70 km s$^{-1}$ for HH 203 and HH 204,
respectively, directed towards the apex of the bow.

There are at least two important questions related to these objects: are HH 203 and HH 204 parts of the same object? and, why does HH 204 show an asymmetry in its brightness distribution?.

Our kinematical results show that:

-- HH 204 has a conical shape (resembling a bow shock) better appreciated at extreme heliocentric velocities: at $V_{helio}$ = --50 km s$^{-1}$ and $V_{helio}$ = +45 km s$^{-1}$.

-- The apex of HH 204 has a complex velocity profile showing a splitting of the main velocity components (at $V_{helio}$ = --24 and +20 km s$^{-1}$) and a blueshifted wing at $V_{helio}$ = --120 km s$^{-1}$. Thus, the apex of HH 204 has the maximum blueshifted velocity of the region, in agreement with the predictions of bow shock models.

-- HH 203 is more jet-like and it seems to be a different entity colliding with a lateral wall of HH 204.

-- We distinguish a pronounced asymmetry in brightness between the bow side of HH 204 near to  the star $\Theta^{2}$ A and the side located away from this star. Henney   (1996) has proposed that a transverse density gradient in the ambient medium where a bow shock propagates, could lead to an asymmetry in brightness of the bow shock. We find some evidence of a transverse density gradient because we find that there is a slight velocity gradient running perpendicular to the axis of HH 204 in the sense that the
fainter regions have larger velocities.

-- HH 203 and HH 204 seem to be part of a structure of large dimensions or lobe. Indeed, Figure 3  is a close-up  of the [NII] velocity map at 
$V_{helio}$ = --14 km s$^{-1}$  showing the field around HH 203 and HH 204, which are located in the SE of this figure. A careful inspection of this figure suggests the detection of an incomplete lobe ending in HH 204 on one side, and at the mark shown in Figure 3 on the other side. This lobe is 132" or 0.29 pc long and it is  more intense in its northern half. The possible existence of this lobe is also revealed in  O'Dell's image of the Orion Nebula published in the National Geographic Supplement (Grosvenor et al. 1995).

\section{ Discussion}
Our results show that HH 202 seems to be part of a larger scale lobe. This lobe is blueshifted relative to the main HII region velocity and shows the large internal motions characteristic of HH flows.  The [NII] velocity cubes show that this lobe (NW lobe) is formed by two `fingers' starting from a region close to the E-W jet discovered by O'Dell  {\it {et al}}. (1997a). HH 203 and HH 204 also seem to be part of a large structure or lobe (SE lobe) of similar dimensions to the ones of the NW lobe. HH 204 is at one of the ends while the other end is located close to the E-W jet, as in the case of the NW lobe. Furthermore, the SE lobe shares the same orientation as the NW lobe. Considering the results presented in the previous sections, we suggest that HH 202 and HH 203-204, are part of a large bipolar outflow, 0.55 pc long, that arises from an object close to the E-W jet. Besides, a preliminary analysis suggests that the NW lobe is blueshifted relative to the background nebula while the SE lobe is redshifted. The measured proper motions of HH 203-204 agree with this interpretation while the proper motions measured a long time ago for HH 202 (Herbig \& Jones 1981) do not support this idea.
On the other hand, the region close to the E-W jet is so rich in objects that it is difficult to identify, by means of the existing stellar data, the object that could be the source of this suggested bipolar outflow. Figure 4 shows the H$\alpha$ velocity map at $V_{helio}$ = -- 50 km s$^{-1}$ of the inner 5' of the Orion Nebula which allows us to have a global view of both lobes. At blueshifted velocities the NW lobe is better detected than the SE lobe. However, jet-like features (such as HH 203 and the `fingers' inside the NW lobe) are detected inside these lobes. Further studies of the stellar content close to the E-W jet and of proper motions of the HH 202 knots would be quite interesting for confirming or rejecting this suggestion. 
 
\section*{Acknowledgements}
The authors wish to thank to Alex Raga for first suggesting the idea of
studying the kinematics of the Orion Nebula with the PUMA equipment.  They
also wish to acknowledge the financial support from grants 400354-5-2398PE of
CONACYT and  IN104696 of DGAPA-UNAM and from scholarships 124449 (CONACYT) and
DGEP-UNAM.  The authors also  thank to Alfredo D\'{\i}az and Carmelo Guzm\'an for
the computer help.  

\clearpage

\vfill\eject

\clearpage

\vfill\eject

\figcaption{ H$\alpha$ velocity map at  $V_{helio}$ = --127 km
s$^{-1}$  of the Orion Nebula obtained with the PUMA FP observations. Some of the important stars, HH objects and proplyds are marked. \label{fig1}}

\figcaption{  Close-up  of the [NII]($\lambda$ 6583 \AA) velocity map at 
$V_{helio}$ = -- 90 km s$^{-1}$  showing the field near to HH 202 which is
located at the NW corner. The arrow to the SE corresponds to one of the ends of the lobe.
\label{fig2}}

\figcaption{ Close-up  of the [NII]($\lambda$ 6583 \AA) velocity map at 
$V_{helio}$ = -- 14 km s$^{-1}$  showing the field near to HH 203 and HH 204 which are located near the SE corner. The mark corresponds to one of the ends of the lobe and it is the same mark than in Figure 2.
 \label{fig3}}

\figcaption{  H$\alpha$ velocity map, at $V_{helio}$ = -- 50 km s$^{-1}$, of the inner 5' of the Orion Nebula. At this blueshifted velocity, the NW lobe, the E-W jet and the jet-like appearance of HH 203 are appreciated. The nebulosity to the W of the E-W jet (marked with an arrow in Figures 2 and 3) corresponds to the location of the possible exciting source of the bipolar flow.
\label{fig4}}

\end{document}